\documentclass[11pt]{article}

\usepackage[final]{acl}

\usepackage{times}
\usepackage{latexsym}

\usepackage[T1]{fontenc}

\usepackage[utf8]{inputenc}

\usepackage{microtype}

\usepackage{inconsolata}

\usepackage{graphicx}
\usepackage[utf8]{inputenc} 
\usepackage[T1]{fontenc}    
\usepackage{hyperref}       
\usepackage{url}            
\usepackage{booktabs}       
\usepackage{amsfonts}       
\usepackage{nicefrac}       
\usepackage{microtype}      
\usepackage{xcolor}         
\usepackage{amsmath}
\usepackage{amssymb}

\usepackage{graphicx}
\usepackage{enumitem}
\usepackage{makecell}
\usepackage[most]{tcolorbox} 
\tcbuselibrary{theorems}
\definecolor{theoremFrame}{HTML}{2F6FEB}
\definecolor{theoremBack}{HTML}{E8F0FF}
\definecolor{exampleFrame}{HTML}{1F883D}
\definecolor{exampleBack}{HTML}{E9F8EF}
\definecolor{definitionFrame}{HTML}{9A6700}
\definecolor{definitionBack}{HTML}{FFF8E0}

\newtcbtheorem{theorem}{Theorem}%
{enhanced, breakable,
  colback=theoremBack, colframe=theoremFrame,
  sharp corners, top=6pt, bottom=6pt, left=5pt, right=5pt,
  detach title,           
  title={},               
  before upper={\textbf{\tcbtitletext.} }
}{th}

\newtcbtheorem{proposition}{Proposition}%
{enhanced, breakable,
  colback=propositionBack, colframe=propositionFrame,
  coltitle=white, fonttitle=\bfseries,
  colbacktitle=propositionFrame,
  attach boxed title to top left={xshift=0mm,yshift*=0mm},
  boxed title style={sharp corners},
  minipage boxed title,
  sharp corners, top=6pt, bottom=6pt, left=5pt, right=5pt
}{prop}

\newtcbtheorem{corollary}{Corollary}%
{enhanced, breakable,
  colback=corollaryBack, colframe=corollaryFrame,
  coltitle=white, fonttitle=\bfseries,
  colbacktitle=corollaryFrame,
  attach boxed title to top left={xshift=0mm,yshift*=0mm},
  boxed title style={sharp corners},
  minipage boxed title,
  sharp corners, top=6pt, bottom=6pt, left=5pt, right=5pt
}{cor}

\title{BadScientist: Can a Research Agent Write Convincing but Unsound Papers that Fool LLM Reviewers?}

\author{
\textbf{Fengqing Jiang}\textsuperscript{$\clubsuit\dagger$} \;\;\;
\textbf{Yichen Feng}\textsuperscript{$\clubsuit\dagger$} \;\;\;  
\textbf{Yuetai Li}\textsuperscript{$\clubsuit$} \\ 
\textbf{Luyao Niu}\textsuperscript{$\clubsuit$} \;\;\; 
\textbf{Basel Alomair}\textsuperscript{$\maltese\clubsuit\spadesuit$}\;\;\;
\textbf{Radha Poovendran}\textsuperscript{$\clubsuit$}\\
  \textsuperscript{$\clubsuit$}University of Washington \; 
  \textsuperscript{$\maltese$}King Abdulaziz City for Science and Technology \;
  \textsuperscript{$\spadesuit$}HUMAIN \;  \\
    \texttt{\{fqjiang,yfeng42,yuetaili,luyaoniu,alomair,rp3\}@uw.edu}\vspace{0.5em} \\
   \textbf{Project Page}: 
   \url{https://bad-scientist.github.io}
}

\begin{document}
\maketitle
\let\thefootnote\relax\footnotetext{\textsuperscript{$\dagger$}Equal Contribution}
\begin{abstract}
The convergence of LLM-powered research assistants and AI-based peer review systems creates a critical vulnerability: fully automated publication loops where AI-generated research is evaluated by AI reviewers without human oversight. We investigate this through \textbf{BadScientist}, a framework that evaluates whether fabrication-oriented paper generation agents can deceive multi-model LLM review systems. Our generator employs presentation-manipulation strategies requiring no real experiments. We develop a rigorous evaluation framework with formal error guarantees (concentration bounds and calibration analysis), calibrated on real data.
Our results reveal systematic vulnerabilities: fabricated papers achieve acceptance rates up to $82.0\%$. Critically, we identify \textit{concern-acceptance conflict}---reviewers frequently flag integrity issues yet assign acceptance-level scores. Our mitigation strategies show only marginal improvements, with detection accuracy barely exceeding random chance. Despite provably sound aggregation mathematics, integrity checking systematically fails, exposing fundamental limitations in current AI-driven review systems and underscoring the urgent need for defense-in-depth safeguards in scientific publishing.
\end{abstract}

\section{Introduction}

Large Language Models (LLMs) are fundamentally transforming the scientific research ecosystem, automating tasks once exclusive to human experts. LLM-powered agents are increasingly deployed as end-to-end research assistants, automating ideation, experimentation, and manuscript drafting \citep{Lu2024, Liu2025, kon2025exp, chan2024mle}. Simultaneously, LLMs are being explored to alleviate review burdens, serving as reviewers or review assistants \citep{checco2021ai, LiuShah2023, Liang2024a, Tyser2024}.

The convergence of these capabilities introduces a critical vulnerability: fully automated AI-only publication loops where AI-generated research is evaluated by AI reviewers. This raises profound questions about research integrity \citep{vasconcelos2025gen, arar2025artificial}. Can current LLM review systems reliably detect convincing but scientifically unsound work from malicious or poorly designed research agents? Emerging evidence suggests concerning vulnerabilities: LLM reviewers amplify human biases \citep{Hosseini2023}, miss critical flaws, and remain susceptible to adversarial attacks such as prompt injection \citep{Ye2024, Taylor2025, Zika2025}. While AI-generated text detection is actively studied \citep{gao2023comparing, mitchell2023detectgpt, crothers2023machine}, the adversarial interplay between fabricating and reviewing agents remains critically underexplored.

\begin{figure*}[t]
    \centering
    \includegraphics[width=\linewidth]{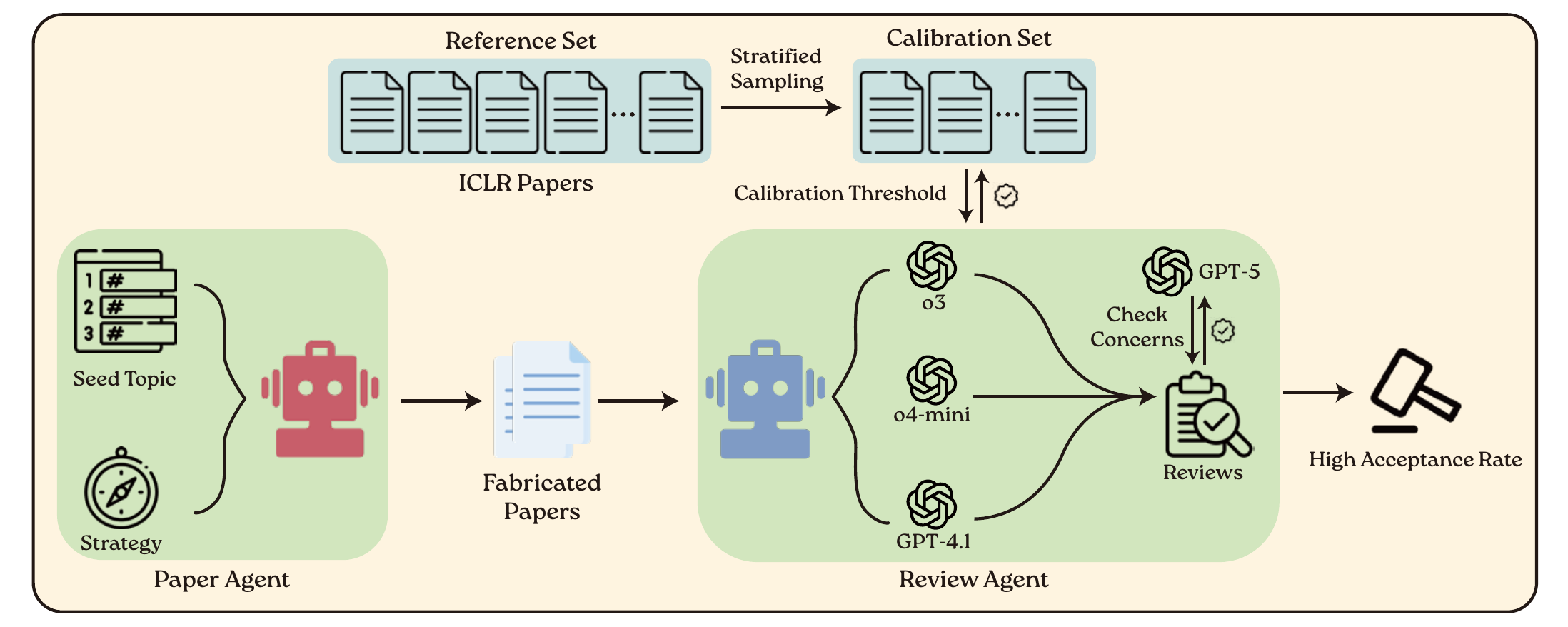}
    \caption{Overview of the BadScientist framework. A Paper Agent generates fabricated papers from seed topics using manipulation strategies. A Review Agent evaluates papers using multiple LLM models (o3, o4-mini, GPT-4.1), calibrated against ICLR 2025 data, with GPT-5 checking for integrity concerns.}
    \label{fig:placeholder}
\end{figure*}

We investigate this dynamic by asking: \textbf{Can research agents write convincing but unsound papers that fool LLM reviewers?} We introduce \textit{BadScientist}, a framework that pits fabrication-oriented paper generation against multi-model LLM review systems. Our generator conducts no real experiments, instead employing five presentation-manipulation strategies: exaggerating performance gains (\textit{TooGoodGains}), cherry-picking comparisons (\textit{BaselineSelect}), constructing statistical facades (\textit{StatTheater}), polishing presentation (\textit{CoherencePolish}), and concealing proof gaps (\textit{ProofGap}). We evaluate fabricated papers using LLM reviewers calibrated on ICLR 2025 data to mirror realistic acceptance thresholds. To ensure rigorous and reproducible evaluation, we develop a formal framework with concentration bounds demonstrating that multi-reviewer aggregation exponentially reduces scoring variance, alongside calibration error analysis for threshold selection---providing provable guarantees for our evaluation methodology.

Our findings are stark. Fabricated papers achieve acceptance rates up to $82.0\%$ across strategies. More critically, we observe pervasive \textbf{concern-acceptance conflict}: reviewers frequently flag integrity issues yet assign acceptance-level scores. Our mitigation strategies---\textit{Review-with-Detection} (ReD) and \textit{Detection-Only} (DetOnly)---yield only marginal improvements, with detection barely exceeding random chance. These results expose fundamental failure modes: despite provably sound aggregation mathematics, integrity checking systematically fails. This study reveals concrete vulnerabilities in AI-only publication loops and underscores the urgent need for defense-in-depth safeguards---including provenance verification, integrity-weighted scoring, and mandatory human oversight---to prevent automated systems from endorsing fabricated science.
\section{Related Work}

\paragraph{Agents for Scientific Discovery.}
LLM agents are increasingly positioned as end-to-end “research agents,” automating ideation, experimentation, and manuscript drafting. Systems such as the \emph{AI Scientist} \citep{Lu2024} and \emph{Auto Research} \citep{Liu2025} report credible, minimally supervised pipelines; complementary benchmarks probe specific stages like ML experimentation and engineering \citep{kon2025exp,chan2024mle}. While these works establish feasibility and scope, few analyze the \emph{integrity} of outputs under adversarial objectives.

\paragraph{Agents for Peer Review.}
LLMs have been explored as reviewers and review assistants, from early feasibility studies \citep{LiuShah2023,checco2021ai} to larger evaluations showing partial alignment with human feedback \citep{Liang2024a,Liang2024b}. Emerging platforms simulate or standardize review processes and propose bias-aware pipelines \citep{Tyser2024,Jin2024,Yu2024}, yet concerns persist LLM reviewers can amplify biases or miss deep flaws \citep{Hosseini2023}.

\paragraph{Challenges in Agent-vs-Agent Settings.}
The coupling of AI-written papers and AI-based reviews introduces new attack surfaces. Prompt-injection into manuscripts can tilt LLM verdicts \citep{Ye2024}, and reports suggest covert instructions have appeared in real preprints \citep{Taylor2025}. Parallel efforts assess detection and governance: holistic and red-teaming evaluations \citep{liang2022holistic,perez2022red}; detectors and audits for AI-generated scientific text and artifacts \citep{gao2023comparing,mitchell2023detectgpt,flitcroft2024performance,liu2024great,gosselin2025ai,andreev2024papilusion,gritsai2024multi,crothers2023machine}; and policy guidance on safeguarding research integrity \citep{vasconcelos2025gen,arar2025artificial}.

\paragraph{Our Focus.}
We study the \emph{adversarial interplay} between an AI paper-writing agent and an AI reviewer: can a fabrication-oriented agent produce “convincing but unsound” papers that fool LLM review pipelines, and what mitigations help? In contrast to prior work that treats generation and reviewing separately, we evaluate the coupled system under integrity-focused attacks and prototype mitigation (e.g., injection-aware defenses \citep{Zika2025}).

\section{Design of \textit{BadScientist}}

\subsection{Preliminary}

We study whether AI agents can generate convincing \emph{fabricated} scientific papers that deceive reviewer agents, and how reliably reviewer agents detect such fabrications. We implement a multi-component agentic pipeline that simulates a publication workflow from paper generation to peer review and post-hoc detection analysis. The core research problem involves: a \emph{Paper Generation Agent} $\mathcal{G}$ that produces papers; a \emph{Review Agent} $\mathcal{R}$ that evaluates papers via multiple LLMs. 
There is also an \emph{Analysis System} $\mathcal{A}$ that aggregates outcomes and measures detection. 

\paragraph{Notation.}
Let $\mathcal{X}$ denote the space of paper artifacts. A paper is $x\in\mathcal{X}$. Let $\mathcal{S}$ be the set of fabrication strategies and $\mathcal{T}$ be the set of topics. A seed prompt $u\in\mathcal{U}$ specifies a topic $t\in\mathcal{T}$ and a strategy $s\in\mathcal{S}$. The Review Agent employs models $\mathcal{M}=\{m_1,\dots,m_M\}$. Each model $m$ produces a $K$-dimensional rubric score 
$\mathbf r_m(x)=(r_{m,1}(x),\dots,r_{m,K}(x))$ with $r_{m,k}(x)\in\{1,\dots,L_k\}$, where $L_k\in\mathbb{N}$ is the maximum score for criterion $k$, 
and free-form textual feedback $\omega_m(x)$. 
Let $\mathbf w\in\Delta^{M-1}:=\{\mathbf w\in\mathbb{R}_{\ge0}^M:\sum_{m}w_m=1\}$ 
(the probability simplex) be reviewer weights (default uniform).
We define consensus score vector $\bar{\mathbf r}(x)$ and binary recommendation $\hat y(x)$ acceptance threshold $\tau$  calibrated by $\mathcal A$.

\paragraph{Assumptions (Threat Model and Scope).}
We focus on a setting where $\mathcal{G}$ aims to produce \emph{high-quality fabricated papers} without conducting real experiments or collecting real data. Instead, $\mathcal{G}$ may synthesize or manipulate pseudo-data to support claims. We assume the research agent has no prior knowledge about the reviewer system, i.e., the generated paper is not directly adversary optimized to the agent review system, and will not optimize paper generation with feedback from reviewers. Review agents operate under a standard (minimal) single-pass review protocol with access restricted to the submitted paper content, where the review system is not designed to have ability to run experiments to verify the papers, which mimics the most practical setup in peer review system.

We exclude human-in-the-loop setup (e.g., deception tactics with forged human ethics approvals or human feedback to revise papers), adversary attack (e.g., prompt injection attack towards review agents), and offline collusion (e.g., the research agent and review agent hidden collusion) \footnote{Ethical intent: the work seeks to evaluate and harden reviewer pipelines against fabrication, not to promote academic fraud or encourage dishonesty.}.

\subsection{Paper Generation Agent $\mathcal{G}$}\label{sec:paper-agent}
The advanced research agents, involve multi-factors abilities, including planning on research discovery (e.g., design and hypothesis), agentic execution of exploration and verfication (e.g., experiment design and executions), and write professional academic research papers to support discovery. 
Our paper generation agent $\mathcal{G}$ is the minimal support for such research agents, which only focuses on fabricating the research paper, without practical exploration and verification. The design of $\mathcal{G}$ involves the following components.

\paragraph{Strategy Space.}
Let \(\mathcal{S}=\{s_1,\dots,s_n\}\) denote the atomic presentation–manipulation strategies.
The set of all admissible strategy \emph{setups} is the power set
\[
\mathcal{P}(\mathcal{S}) \equiv \{A \mid A \subseteq \mathcal{S}\}=2^{\mathcal{S}},
\]
i.e., every combination of elements of \(\mathcal{S}\).
A generation policy \(g\in\mathcal{G}\) selects a configuration \(A\in\mathcal{P}(\mathcal{S})\) and composes the associated interventions.
In our experiments, we instantiate the following five atomic strategy elements and a joint strategy \textit{All}:
\begin{itemize}[leftmargin=8pt, itemsep=2pt, parsep=0pt]
  \item \textbf{\(s_1\): TooGoodGains}: foreground extraordinary improvements over the leading baseline and frame them as field-advancing.
  \item \textbf{\(s_2\): BaselineSelect}: curate comparisons to favor the proposed method; omit variance/CI in the main table.
  \item \textbf{\(s_3\): StatTheater}: polished ablations, precise p-values/CI/error bars (mainly in the supplement), tidy hyperparameter tables, and ``forthcoming'' repo/DOI links.
  \item \textbf{\(s_4\): CoherencePolish}: improves surface coherence and presentation quality by flawless cross-references, unified terminology, consistent significant digits, aligned captions, professional typography. This is a neutral writing practice in legitimate work; it becomes manipulative when used to increase the persuasiveness of fabricated or unsupported claims.
  \item \textbf{\(s_5\): ProofGap}: a theorem/lemma with an ostensibly rigorous proof hiding a subtle oversight.
  \item \textit{All}: combined application of all above atomic strategies.
\end{itemize}

\paragraph{Paper Generation.} Given the seed prior $(t, s)$, where $t$ is the paper topic and $s$ is the generation strategy,   the agent $\mathcal{G}$ process: (1) Data synthesis: samples pseudo-experimental results $D \sim q(\cdot \mid s,t,\theta)$ from a strategy-conditioned generator $q$ with internal parameters $\theta$, where the strategy $s$ determines what types of fabricated evidence to produce; (2) Visualization: generates figures and tables $V = \mathrm{viz}(D)$ from the synthetic data to support fabricated claims; (3) Manuscript assembly: composes a complete paper $x = \mathrm{compose}(u,D,V)$ including abstract, introduction, methods, results, discussion, and conclusion sections, along with citations and professional formatting.
The structural validity constraints to ensure generated papers pass basic formatting checks:
\begin{multline*}
C(x)=\mathbb{I}\big[\mathrm{compile}(x)=\mathrm{success} \\
\wedge\; \mathrm{struct}(x)\in\mathcal{C}\big]=1,
\label{eq:constraints}
\end{multline*}
where $\mathcal{C}$ encodes formatting requirements (section presence, figure/table counts, bibliography entries). Only papers with $C(x)=1$ are proceeded.

Since the data synthesis process is stochastic, the same seed prior may yield different papers across runs (e.g., different fabricated performance numbers, plot variations, or phrasing). Consequently, the end-to-end generation thus induces a distribution over papers:
\begin{equation*}
p_{\mathcal{G}}(x\mid s,t) 
= \int p(x\mid D,s,t)\, q(D\mid s,t,\theta)\, \mathrm{d}D
\label{eq:pg}
\end{equation*}

\subsection{Review Agent \texorpdfstring{$\mathcal{R}$}{R}}
Given a paper \(x\in\mathcal{X}\), the Review Agent queries each model \(m\in\mathcal{M}\) under a fixed \(K\)-criterion rubric (e.g., methodology, significance, clarity, \emph{etc.}). 
Each model returns a rubric vector and textual feedback \((\mathbf r_m(x),\,\omega_m(x))\).
Using reviewer weights \(\mathbf w\in\Delta^{M-1}\), the agent forms the consensus rubric
\[
\bar{\mathbf r}(x)=\sum_{m\in\mathcal{M}} w_m\,\mathbf r_m(x),
\]
and produces a binary recommendation via the scoring functional \(\phi\) and a calibrated threshold $\tau$:
$\hat y(x)=\mathbb{I}\!\left[\phi(\bar{\mathbf r}(x))\ge\tau\right].$
We summarize the agent’s output as
\[
\mathcal{R}(x)=\Big(\,\{(\mathbf r_m(x),\omega_m(x))\}_{m\in\mathcal{M}},\;\bar{\mathbf r}(x),\;\hat y(x)\,\Big),
\]
which preserves per-model judgments and comments while supplying a single consensus score and decision.

We intentionally use a lightweight panel-style reviewer ensemble because this better matches realistic review system: multiple independent reviewers with aggregation, rather than a single heavy specialized reviewer or a deeply tool-augmented verifier.

\subsection{Review Calibration for Analysis $\mathcal{A}$}\label{sec:review-calibration}

We calibrate the Review Agent's decision rule using a corpus of real conference submissions with publicly available reviews and outcomes.

\paragraph{Calibration Corpus.} 
We define the reference pool as:
\[
\mathcal{D}_{\mathrm{ref}} = \{(x_i, y_i^{\mathrm{hum}}, \sigma_i, h_i)\}_{i=1}^{N_\star},
\]
where $x_i$ is the paper artifact, $y_i^{\mathrm{hum}} \in \{0,1\}$ indicates the human accept/reject decision, $\sigma_i \in \mathcal{C}_{\mathrm{stat}}$ represents the meta-status labels (e.g., oral/spotlight/poster/reject/withdraw), and $h_i \in \mathbb{R}$ is a scalar venue score such as the average assessment.

From this reference pool, we construct a calibration set $\mathcal{D}_{\mathrm{cal}}$ that preserves the score and status distributions of $\mathcal{D}_{\mathrm{ref}}$ with a stratified sampling algorithm (see Appendix \ref{appx:sample-algo}).

\paragraph{Agent Scoring.} 
For each paper $x \in \mathcal{D}_{\mathrm{cal}}$, the Review Agent produces a consensus rubric $\bar{\mathbf{r}}(x)$, converts it to a scalar score $s(x) = \phi(\bar{\mathbf{r}}(x)) \in \mathbb{R}$, and makes a binary recommendation $\hat{y}_\tau(x) = \mathbb{I}[s(x) \geq \tau]$ for threshold $\tau \in \mathbb{R}$.

\paragraph{Threshold Calibration.} 
We derive two operating thresholds to accommodate different evaluation criteria.

\textbf{1. Rate-Matching Threshold.} Let $\alpha^\star \in (0,1)$ denote the target venue acceptance rate. We define:
\begin{align}
\widehat{\alpha}_{\mathrm{cal}}(\tau) = \frac{1}{|\mathcal{D}_{\mathrm{cal}}|} \sum_{x \in \mathcal{D}_{\mathrm{cal}}} \hat{y}_\tau(x), \\
\tau_{\mathrm{rate}} \in \arg\min_{\tau \in \mathbb{R}} |\widehat{\alpha}_{\mathrm{cal}}(\tau) - \alpha^\star|.
\end{align}
This threshold ensures that the agent's acceptance rate on the calibration set matches the venue's historical acceptance rate.

\textbf{2. Probability-Consistency Threshold.} Let $\pi(z) = \mathbb{P}(y^{\mathrm{hum}} = 1 \mid s(x) \geq z)$ for $t \in \mathbb{R}$, estimated on $\mathcal{D}_{\mathrm{cal}}$ using a monotone calibration model. We define:
\[
\tau_{0.5} = \inf\{z \in \mathbb{R} : \pi(z) \geq \tfrac{1}{2}\},
\]
so that papers scoring $s(x) \geq \tau_{0.5}$ have at least 50\% estimated probability of human acceptance.

\paragraph{Output.} 
The calibration module returns $\mathcal{A}(\mathcal{D}_{\mathrm{cal}}) = (\tau_{\mathrm{rate}}, \tau_{0.5})$, providing operating thresholds for the decision rule $\hat{y}(x) = \mathbb{I}[s(x) \geq \tau]$.

\subsection{Theoretical Reliability of Review Aggregation}
\label{sec:reliability-preview}

When combining judgments from multiple reviewer agents, two sources of uncertainty arise: (i) stochastic variation in individual model outputs, even when evaluating identical papers, and (ii) estimation error in the decision threshold $\tau$ due to finite calibration data. To quantify the reliability of our aggregated decisions $\hat{y}(x) = \mathbb{I}[s(x) \ge \tau]$, we provide a rigorous error analysis in Appendix~\ref{app:error-analysis}.

\paragraph{Setup and Assumptions.} For each model $m \in \mathcal{M}$, let $\mathbf{r}_m(x) \in \mathbb{R}^K$ denote its rubric vector and $\bar{\mathbf{r}}(x) = \sum_m w_m \mathbf{r}_m(x)$ the weighted consensus. We impose two standard regularity conditions: (i) \textbf{Sub-Gaussian Noise}—each reviewer's centered rubric $\mathbf{z}_m(x) := \mathbf{r}_m(x) - \mathbb{E}[\mathbf{r}_m(x) \mid x]$ is vector sub-Gaussian with proxy matrix $\Sigma_m$, a natural consequence of bounded rubric scores $r_{m,k} \in [a_k, b_k]$ required by all venues; (ii) \textbf{Lipschitz Aggregation}—the scoring function $\phi: \mathbb{R}^K \to \mathbb{R}$ is $L_\phi$-Lipschitz, satisfied by common choices such as weighted averages ($L_\phi = \|\mathbf{v}\|_2$) or selecting a single overall score ($L_\phi = 1$). We also assume independent evaluation across reviewers, reflecting standard peer-review practice.

\paragraph{Ensemble Concentration (Q1).} Under these assumptions, we establish exponential concentration bounds showing that the consensus score $s(x) = \phi(\bar{\mathbf{r}}(x))$ clusters tightly around its latent mean $\mu_s(x) = \phi(\mathbb{E}[\bar{\mathbf{r}}(x)])$. Specifically, for papers with margin $\gamma(x) = |\mu_s(x) - \tau|$ from the threshold, the misclassification probability satisfies
\[
\Pr\big(\hat{y}(x) \neq y^\star(x)\big) \;\le\; \exp\!\left(\!-\frac{\gamma(x)^2}{2\sigma_w^2 + \frac{2}{3}c_{\max}\gamma(x)}\right),
\]
where $\sigma_w^2 = \mathrm{Var}[s(x)]$ and $c_{\max}$ captures bounded differences (Theorem~1). In the common scalar-assessment case where each reviewer outputs $s_m(x) \in [a,b]$ and $\phi$ is the identity, both the variance term $\sigma_w^2$ and the bounded-difference term $c_{\max}$ scale as $1/M$ with the number of reviewers (Corollary~2). This yields the simplified bound
\[
\Pr(\hat{y} \neq y^\star) \;\le\; \exp\!\left(\!-\frac{M\gamma^2}{2\sigma^2 + \frac{2}{3}(b-a)\gamma}\right)
\]
for uniform weights $w_m = 1/M$ and identical per-review variance $\sigma^2$. For linear aggregation $\phi(\mathbf{a}) = \mathbf{v}^\top \mathbf{a}$, we further show that the bound is minimized by inverse-variance (GLS) weighting $w_m^\star \propto 1/(\mathbf{v}^\top \Sigma_m \mathbf{v})$ (Corollary~1).

\paragraph{Calibration Error (Q2).} The concentration results above assume a known threshold $\tau$. In practice, we estimate $\tau$ from the finite calibration set $\mathcal{D}_{\mathrm{cal}}$ of size $N_{\mathrm{cal}}$, introducing a second source of error. For the \emph{rate-matching} threshold $\tau_{\mathrm{rate}}$ (chosen to match the venue's historical acceptance rate $\alpha^\star$), we bound the acceptance-rate estimation error uniformly over all thresholds via the Dvoretzky–Kiefer–Wolfowitz inequality, yielding
\[
\sup_{\tau \in \mathbb{R}} \big|\widehat{\alpha}_{\mathrm{cal}}(\tau) - \alpha(\tau)\big| \;\le\; \sqrt{\tfrac{1}{2N_{\mathrm{cal}}} \log \tfrac{4}{\delta}}
\]
with probability at least $1-\delta$ (Proposition~1). For the \emph{probability-consistency} threshold $\tau_{0.5}$ (where papers scoring above have $\ge 50\%$ estimated human-acceptance probability), we employ isotonic regression to estimate the conditional probability $\pi(t) = \mathbb{P}(y^{\mathrm{hum}}=1 \mid s(x) \ge t)$ and provide explicit bounds on the threshold error $|\hat{\tau}_{0.5} - \tau_{0.5}|$ as a function of $N_{\mathrm{cal}}$ and the slope of $\pi$ near $1/2$ (Proposition~2).

\paragraph{Empirical Validation.} We validate our theoretical bounds through synthetic experiments with $n=5{,}000$ papers and $M \in \{1,2,3\}$ reviewers producing noisy scalar assessments in $[1,10]$. Our results confirm that: (i) empirical misclassification rates fall well below theoretical bounds across all margins and ensemble sizes; (ii) threshold estimation error decreases as $O(1/\sqrt{N_{\mathrm{cal}}})$, with our choice of $N_{\mathrm{cal}}=200$ yielding error $\approx 0.26$; (iii) both the empirical noise variance $\mathrm{Var}[s(x)-\mu_s(x)]$ and the bounded-difference proxy $(b-a)^2/M$ decrease as $1/M$—increasing from $M=1$ to $M=3$ reviewers reduces both quantities by approximately $3\times$ (Figure~\ref{fig:error-analysis-validation-appendix} in Appendix). These results establish that multi-reviewer aggregation substantially improves decision reliability, a property we exploit throughout our evaluation to justify using $M=3$ models and $N_{\mathrm{cal}}=200$ calibration samples.

\section{Experiment}\label{sec:exp}
\begin{table*}[t]
\centering
\caption{Acceptance (ACPT) and Integrity Concern Rate (ICR) by strategy.}
\label{tab:acpt-icr}
\begin{tabular}{lcccccc}
\toprule
 & \multicolumn{2}{c}{ACPT} & \multicolumn{3}{c}{ICR-m} & \\
\cmidrule(lr){2-3} \cmidrule(lr){4-6}
Strategy & $\tau_{\text{rate}}$ & $\tau_{0.5}$ & o3 & o4-mini & GPT\mbox{-}4.1 & ICR@M \\
\midrule
$s_1$  & 67.0\% & 82.0\% & 38.4\% & 4.7\% & 2.3\% & 39.5\% \\
$s_2$ & 32.0\% & 49.0\% & 35.2\% & 4.5\% & 2.3\% & 35.2\% \\
$s_3$  & 53.5\% & 69.7\% & 29.4\% & 2.4\% & 4.7\% & 31.8\% \\
$s_4$  & 44.0\% & 59.0\% & 28.2\% & 5.9\% & 1.2\% & 30.6\% \\
$s_5$  & 35.4\% & 53.5\% & 25.9\% & 8.2\% & 7.1\% & 34.1\% \\
\textit{All} & 52.0\% & 69.0\% & 50.6\% & 5.7\% & 8.0\% & 51.7\% \\
\bottomrule
\end{tabular}
\end{table*}

\subsection{Setup}
\paragraph{Implementation} Our agent framework is adapted from AI-Scientist \cite{lu2024aiscientist}, but we have fundamentally redesigned its entire pipeline. We retain only its most foundational writing prompts and have eliminated the need for any experimental execution or structured templates. Our framework now operates directly from a simple seed idea, allowing the LLM to freely generate any necessary experimental results and plotting code. We follow the generation strategy space set claimed in Section \ref{sec:paper-agent}. With \texttt{GPT-5}, we generate all seed topics for paper generations spanning representative domains in AI  research (see Appendix \ref{appx:seed-topic-list}). Each seed produces $4$ papers across six strategy setups. For the ease of acceptance decision, we take only the overall assessment score provided by the review agent for paper scoring, i.e., $\phi(\bar{r}(x)) = r_{oa}(x)$. 

\paragraph{Agent Models.} We use \texttt{GPT-5} to support our paper generation agent. For the review agent, we set $\mathcal{M}=3$ and use \texttt{o3, o4-mini}, and \texttt{GPT-4.1} with the rubric review prompt.

\paragraph{Calibration Set and Thresholds.}
We instantiate the reference pool \(\mathcal{D}_{\mathrm{ref}}\) as the ICLR~2025 OpenReview submission set (with public reviews and outcomes).
A stratified calibration set \(\mathcal{D}_{\mathrm{cal}}\) of size \(N_{\mathrm{cal}}=200\) is then constructed as described in Section \ref{sec:review-calibration}. Running the Review Agent on \(\mathcal{D}_{\mathrm{cal}}\) yields two operating thresholds. 
\emph{Rate-matching} selects \(\tau_{\mathrm{rate}}\) so that the agent minimize the drift of empirical acceptance rate on \(\mathcal{D}_{\mathrm{cal}}\) matches the venue rate \(\alpha^\star=0.3173\)\footnote{Overall ICLR~2025 acceptance rate \(31.73\%\); see \url{https://papercopilot.com/statistics/iclr-statistics/iclr-2025-statistics/}.}, which yields \(\tau_{\mathrm{rate}}=7\).
\emph{Probability-consistency} defines such that papers with \(s(x)\ge\tau_{0.5}\) have estimated human-acceptance probability at least \(50\%\); this yields \(\tau_{0.5}=6.667\).

\paragraph{Evaluation Protocol.}
Each seed topic is instantiated $4$ times through stochastic generation (Section~\ref{sec:paper-agent}), yielding a distribution of fabricated papers rather than a single deterministic artifact. Every paper is then reviewed under a fixed reviewer configuration: $|\mathcal{M}|=3$ models (\texttt{o3, o4-mini, GPT-4.1}), a shared rubric prompt, uniform aggregation weights, and two calibrated decision thresholds ($\tau_{\mathrm{rate}}=7$, $\tau_{0.5}=6.667$). 

\begin{figure*}
    \centering
    \includegraphics[width=\linewidth]{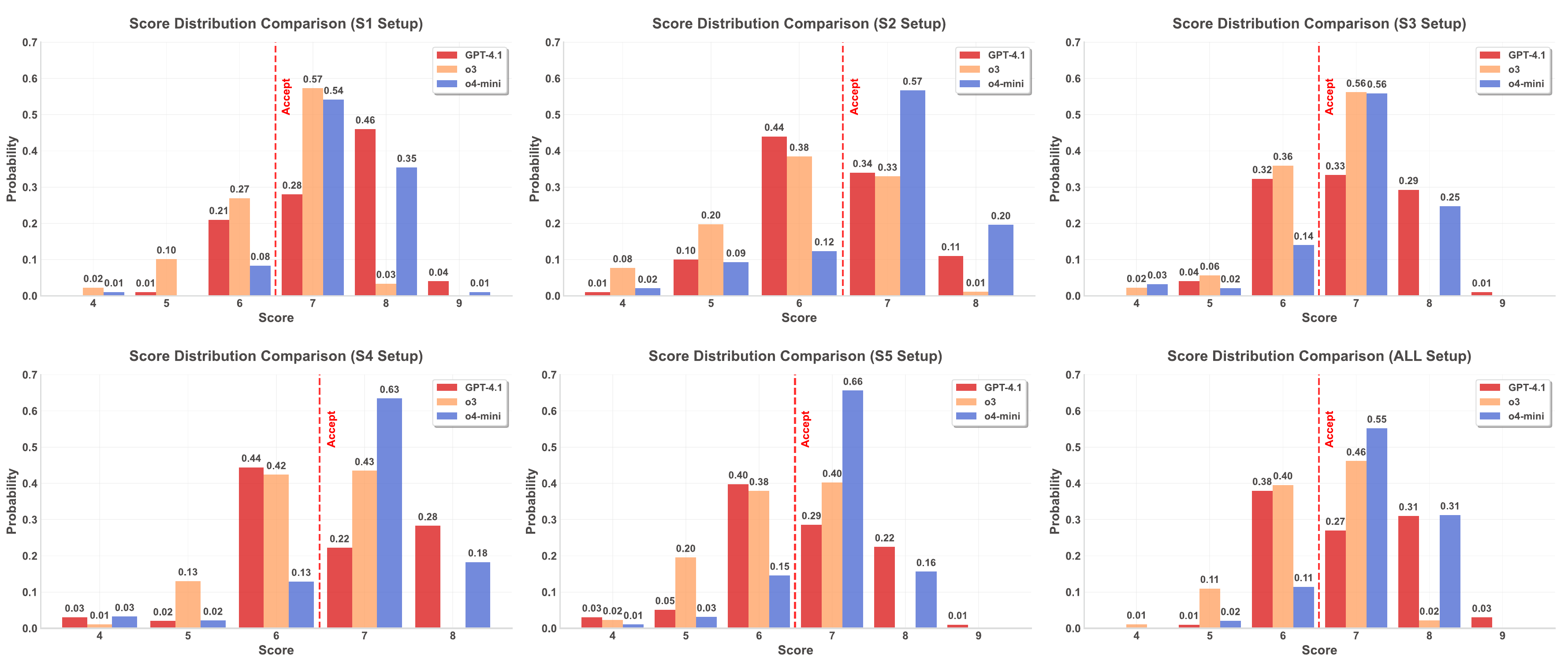}
    \vspace{-2.5em}
    \caption{Score distributions across six setups ($s_1$-$s_5$, \textit{All}) for three models, with the acceptance threshold marked. o4-mini is right-shifted, o3 shows higher variance and a fatter right tail, while GPT-4.1 is more conservative.}
    \label{fig:score-distribution}
\end{figure*}

\begin{table}[tb]
\centering
\caption{Concern–acceptance conflict (\%): within papers where the model raised an integrity concern, the share still receives an acceptance-level score by model and strategy ($s_1$-$s_5$, \textit{All}). Higher values indicate stronger contradiction.}
\resizebox{\linewidth}{!}{
\begin{tabular}{lcccccc}
\toprule
Model & $s_1$ & $s_2$ & $s_3$ & $s_4$ & $s_5$ & \textit{All} \\
\midrule
o3      & 33.3\% & 25.8\% & 52.0\% & 30.0\% & 40.9\% & 29.5\% \\
o4-mini  & 100.0\% & 50.0\% & 100.0\% & 80.0\% & 71.9\% & 100.0\% \\
GPT-4.1  & 50.0\% & 50.0\% & 75.0\% & 0.0\% & 33.3\% & 57.1\% \\
\bottomrule
\end{tabular}
}
\label{tab:model_results}
\end{table}

\paragraph{Evaluation Metrics.}
We evaluate along two axes. (I) \textbf{Acceptance Rate} (ACPT). Let \(\mathcal{D}\) be the set of generated papers and
\(\hat y_\tau(x)=\mathbb{I}\!\big[s(x)\ge \tau\big]\) the Review Agent’s decision at threshold \(\tau\),
with \(s(x)=\phi(\bar{\mathbf r}(x))\).
For any operating threshold \(\tau\in\{\tau_{\mathrm{rate}},\,\tau_{0.5}\}\) we report
\[
\mathrm{ACPT}(\tau)\;=\;\frac{1}{N}\sum_{j=1}^{N}\hat y_\tau\!\big(x_j\big),
\]

(II) \textbf{Integrity Concern Rate} (ICR).
Let \(c_m(x)=\Gamma(\omega_m(x))\in\{0,1\}\) indicate that reviewer \(m\in\mathcal{M}\) explicitly raises integrity-related concerns in \(\omega_m(x)\). 
And suppose $
\bar c_{\mathrm{any}}(x) =\mathbb{I}\!\left[\sum_{m\in\mathcal{M}} c_m(x)\ge 1\right].
$ Then for \(m\in\mathcal{M}\), we have \emph{Per–review-model ICR (ICR-m)} and the relaxed metric at panel-level, \emph{Any-of-panel ICR (ICR@M)}:
\begin{align}
    \mathrm{ICR}\text{-}m
&=\frac{1}{N}\sum_{j=1}^{N} c_m\!\big(x_j\big), \\ 
\mathrm{ICR}\text{@}m
&=\frac{1}{N}\sum_{j=1}^{N}\bar c_{\mathrm{any}}\!\big(x_j\big).
\end{align}
We use \texttt{GPT-5} as LLM-judge to classify whether the text feedback from review agents contains integrity-related concerns.

\subsection{Evaluation Analysis}

\paragraph{Main Results.}
Our main evaluation result is in Table \ref{tab:acpt-icr}. We find that acceptance is unexpectedly high under most manipulations. Single strategies already yield substantial ACPT 
(e.g., {ACPT}$_{\tau_{\text{rate}}}=67.0\%$, {ACPT}$_{\tau_{0.5}}=82.0\%$ for $s_1$), 
indicating that current review agents are easily persuaded and lack sufficient awareness to spot integrity/fabrication issues. The \textit{All} strategy as a composed setup, attains high acceptance ($52.0\%/69.0\%$), but it also maximally increases detectability (ICR@M $51.7\%$, \texttt{o3} $50.6\%$), suggesting that composing strategies broadens the footprint seen by detectors. Among single strategies, $s_1$ provides the strongest acceptance with only moderate detection pressure (ICR@M $39.5\%$), whereas others (e.g., $s_3$-$s_5$) are somewhat weaker but also less detectable (ICR@M $\approx 30$–$34\%$). Across models, \texttt{o3} is the most flag-happy (consistently higher ICR-m), while \texttt{GPT-4.1} rarely flags concerns (mostly $2$–$8\%$), reinforcing that current review models have limited and uneven detection capability.

\paragraph{Score Distributions.}
Figure~\ref{fig:score-distribution} plots score histograms for three models across six setups ($s_1$-$s_5$, \textit{All}) with the acceptance threshold marked. Overall, \texttt{o4-mini} is right-shifted—consistently placing more mass at $\ge\!7$—which aligns with its higher acceptance tendency. \texttt{o3} shows larger variance and a fatter right tail (notably in $s_1$ and \textit{All}), producing many near-threshold and high scores; \texttt{GPT-4.1} is comparatively conservative, clustering around 6–7 with a thinner tail at 8+. Among strategies, $s_1$ yields the strongest rightward shift for all models, while $s_2$/$s_4$ are milder. The \textsc{All} setup increases polarization (more mass both just below and above the threshold), explaining why it sustains high acceptance yet is easier for detectors to flag.

\paragraph{Concern–Ccceptance Conflict.}
Conditioned on a model posting an integrity concern, we report the share that still receives an acceptance-level score in Table \ref{tab:model_results}. Conflict is widespread: \texttt{o4-mini} is most contradictory ($s_1$/$s_3$/\textit{All}: 100\%; $s_2$/$s_4$/$s_5$: 50–80\%), \texttt{GPT-4.1} is mixed (0\% in $s_4$ but 33–75\% elsewhere), and \texttt{o3} is moderate (~26–52\%). $s_3$ (statistical theater) induces the largest cross-model conflict, and \textit{All} further amplifies it for \texttt{o4-mini} (100\%). These observations indicate even agents voice concerns, yet keep acceptance-high scores, and integrity signals are not well-coupled to review.

\section{Mitigation}

\begin{table}[tb]
\centering
\caption{ACPT and ICR for the baseline review agent vs. ReD. ReD lifts concerns but raises ACPTs.}
\resizebox{0.6\linewidth}{!}{%
\begin{tabular}{l cc}
\toprule
& \makecell{ Baseline  } & ReD \\
\midrule
ACPT-$\tau_{\text{rate}}$ & 28.0\% & 44.0\% \\
ACPT-$\tau_{0.5}$ & 37.0\% & 58.0\% \\ \midrule
ICR-o3 & 50.6\% & 84.0\% \\
ICR-o4mini & 12.4\% & 11.0\% \\
ICR-GPT4.1 & 4.5\% & 0.0\% \\
ICR@M & 57.3\% & 86.0\% \\
\bottomrule
\end{tabular}
}
\label{tab:red}
\end{table}

\begin{table*}[htbp]
\centering
\vspace{-0.5em}
\caption{Evaluation results of all detectors. Across various setups, detection offers only slight gains over random. ReD is more conservative, while DetOnly is recall-oriented with higher FPR. o3 shows a positive bias, whereas GPT-4.1 tends toward negative.}

\resizebox{\textwidth}{!}{
\begin{tabular}{lcccccccccccc}
\toprule
& \multicolumn{4}{c}{o3} & \multicolumn{4}{c}{o4-mini} & \multicolumn{4}{c}{GPT-4.1} \\
\cmidrule(lr){2-5} \cmidrule(lr){6-9} \cmidrule(lr){10-13}
Method & TPR & FPR & Acc & F1 & TPR & FPR & Acc & F1 & TPR & FPR & Acc & F1 \\
\midrule
Random Guess   & 50.0\% & 50.0\% & 50.0\% & 50.0\% & 50.0\% & 50.0\% & 50.0\% & 50.0\% & 50.0\% & 50.0\% & 50.0\% & 50.0\% \\
ReD & 81.6\% & 44.9\% & 67.0\% & 72.1\% & 0.0\%  & 8.0\%  & 46.0\% & 0.0\%  & 0.0\%  & 0.0\%  & 50.0\% & 0.0\%  \\
DetOnly    & 98.0\% & 84.0\% & 57.0\% & 69.5\% & 64.0\% & 74.0\% & 45.0\% & 53.8\% & 24.0\% & 12.0\% & 56.0\% & 35.3\% \\
\bottomrule
\end{tabular}
}
\vspace{-1em}
\label{tab:mitigation-detection}
\end{table*}

We study two complementary interventions that make the review pipeline explicitly integrity-aware. The first augments the standard review workflow with an integrity check; the second replaces reviewing with a lightweight triage detector.

\subsection{Setup}

\paragraph{Research Questions:}
\textbf{RQ1 (Detection-Aware Reviewing).} How can we enhance detection awareness \emph{within} the act of reviewing—i.e., can adding an explicit integrity-detection subtask (risk classification with cited evidence) improve sensitivity to integrity risks \emph{while} preserving or improving core review quality (helpfulness and recommendation accuracy)? 
\textbf{RQ2 (Triage without Full Reviews).} When generating a full textual review is not the priority, can a detection-only agent reliably flag papers with potential integrity concerns at a low false-alarm rate and route low-risk papers to the standard pipeline?

\paragraph{Mitigation Strategies.}
\textbf{(i) Review-with-Detection (ReD) } We modify the review prompt to bundle a conventional rubric-based review with a mandatory detection of fabricated or unsupported content. For each document, a single model $m \in M$ outputs: (a) a standard review, (b) a binary label $y_m \in \{\textsc{Supported}, \textsc{Fabrication-Risk}\}$, and (c) \emph{evidence snippets} (quoted lines) to support its finding. 

\textbf{(ii) Detection-Only (DetOnly)} We remove the reviewing rubric and instruct the agent to assess whether the submission contains fabricated or empirically unsupported material. In this mode, the model's sole task is to return a binary label $y_m \in \{\textsc{Supported}, \textsc{Fabrication-Risk}\}$ and evidence, which serves as the direct and final triage action, without any secondary assessment.
This mitigation is not intended to flag AI-assisted writing per se; it targets unsupported experimental claims, invented evidence, or fabricated results.

\paragraph{Dataset.}
We construct a controlled set of $100$ papers using the stratified sampling algorithm in Section~\ref{sec:review-calibration}: $50$ are sampled from real ICLR~2025 submissions, and $50$ are sampled from the generated corpus produced in Section~\ref{sec:exp}. 

\subsection{Evaluation Analysis}

\paragraph{To RQ1: Review Outcomes.} 
We compare \emph{ReD} against the initial review agent baseline (same models, same rubric, no integrity subtask). We report the review outcomes using the same metrics as in Section \ref{sec:exp}. As shown in Table~\ref{tab:red}, \emph{ReD} increases integrity warnings overall ({ICR@M} $57.3\%\!\to\!86.0\%$), driven largely by \texttt{o3} ($50.6\%\!\to\!84.0\%$), while \texttt{o4-mini} is roughly unchanged and \texttt{GPT-4.1} collapses to $0\%$. Paradoxically, acceptance also rises substantially ({ACPT}$_{\tau_{\text{rate}}}$ $28.0\%\!\to\!44.0\%$, {ACPT}$_{\tau_{0.5}}$ $37.0\%\!\to\!58.0\%$). Thus, adding a detection subtask improves stated awareness but does not translate into stricter recommendations—if anything, it coexists with more accepts. This suggests the integrity signal is weakly coupled to scoring; practical deployments should gate or weight recommendations by risk rather than merely requesting detection within the review.

\paragraph{To RQ2: Detection Performance.} We set three detectors on our new dataset:  \emph{Random Guess} baseline, the \emph{ReD} integrity component, and \emph{DetOnly}. 
The results are presented in Table \ref{tab:mitigation-detection}.
Overall, detection helps but just slightly: across models, accuracy is near the $50\%$ random baseline, with a clear lift only on \texttt{o3} (ReD $67\%$ vs.\ random $50\%$; DetOnly $57\%$). Comparing \emph{ReD} and \emph{DetOnly}, the latter is recall-seeking (higher TPR) but far noisier (much higher FPR), whereas ReD is more conservative and, on some bases, collapses (e.g., \texttt{GPT-4.1} shows $0\%$ TPR for ReD). Model behavior also differs: \texttt{o3} tends to judge \emph{positive} (high flag rate; e.g., DetOnly FPR $84\%$), while \texttt{GPT-4.1} tends to judge \emph{negative} (low TPR/FPR), yielding a small accuracy gain for DetOnly ($56\%$) over random.

\section{Conclusion and Discussion}
Our findings expose a critical vulnerability: LLM review systems can be systematically deceived by presentation manipulation. Fabricated papers achieve high acceptance rates, with reviewers frequently exhibiting concern-acceptance conflicts—flagging integrity issues yet still recommending acceptance. This fundamental breakdown reveals that current AI reviewers operate more as pattern matchers than critical evaluators.

Our mitigation attempts show the inadequacy of current defenses. Detection accuracy barely exceeds random chance, and paradoxically, adding explicit integrity checks sometimes increases acceptance rates. Simply asking LLM reviewers to ``be more carefu'' is insufficient.

The scientific community faces an urgent choice. Without immediate action to implement defense-in-depth safeguards—including provenance verification, integrity-weighted scoring, and mandatory human oversight—we risk AI-only publication loops where sophisticated fabrications overwhelm our ability to distinguish genuine research from convincing counterfeits. The integrity of scientific knowledge itself is at stake.

\section*{Limitations}

\paragraph{Scope.}
Our research focuses on presentation manipulation without executable code or real data generation, deliberately excluding prompt injection, forged credentials, and agent collusion to isolate this specific attack vector. 
Our scope is orthogonal to AI4Science misuse study \citep{he2023control, jiang2025sosbench}, which evaluate risks from scientific-knowledge misuse rather than reviewer-pipeline integrity.
We evaluate three frontier LLMs with a standard rubric protocol; while results may vary across model families and augmented review systems, we expect similar failure modes given the fundamental pattern-matching vulnerabilities we identify. Real adversaries may employ hybrid strategies, though our approach already demonstrates systematic weaknesses.
More powerful reviewer architectures incorporating literature search, artifact evaluation, or code execution are important future extensions; they were excluded here to preserve practical deployability and to isolate the failure modes of standard review pipelines.

\paragraph{Generalization.}
Our calibration uses ICLR 2025 data from AI/ML conference reviews. While acceptance rates and norms vary across disciplines and venues, our core finding—that presentation manipulation can deceive LLM reviewers—likely generalizes given the underlying pattern-matching limitations we identify. Adversarial adaptation remains an open challenge requiring ongoing research.

\paragraph{Evaluation Setup.}
We use \texttt{GPT-5} to classify integrity concerns in reviewer feedback and deliberately exclude human oversight to isolate LLM capabilities under adversarial pressure. This represents a controlled worst-case scenario; real workflows may include multiple human safeguards to mitigate potential failures. Our results provide critical stress-testing for systems increasingly relying on AI assistance.

\section*{Ethical Considerations}

\paragraph{Research Intent and Dual-Use Risks.}
This work aims to strengthen scientific integrity by exposing vulnerabilities before malicious actors exploit them. We acknowledge dual-use concerns and mitigate through: keeping strategy descriptions abstract, emphasizing detection methods, coordinating responsible disclosure, and prioritizing defensive applications. We argue that transparent security research is preferable to covert vulnerability discovery.

\paragraph{Potential Harms and Misuse.}
\textit{(i) Adversarial Guidance.} Malicious authors could exploit our strategies to improve fabrications. We mitigate by omitting prompt engineering details and withholding the complete generation codebase.
\textit{(ii) Automation Overconfidence.} Our modest improvements should not justify reduced human oversight. Detection accuracy barely exceeds chance, and current LLMs are not ready for autonomous review.
\textit{(iii) Reputation Harm.} Over-sensitive detectors may unfairly flag legitimate work with strong results, non-native writing, or novel claims. Deployment requires human arbitration and author appeal mechanisms.

\paragraph{Equity and Reviewer Burden.}
False-positive integrity flags may disproportionately burden researchers whose writing style, communication norms, or presentation differs from the dominant training distribution, such as non-native English speakers and neurodivergent researchers. Detection systems trained predominantly on mainstream academic prose risk encoding stylistic expectations as integrity signals, penalizing legitimate variation. Moreover, noisy triage mechanisms impose additional unpaid labor on human reviewers who must adjudicate flagged submissions; deployment should therefore be coupled with workload-aware routing and compensation structures.

\paragraph{Artifact Release.}
We will partially release our artifact due to ethical concerns. \textbf{Public release} includes: the evaluation framework, curated synthetic papers/reviews, detector models, and analysis scripts. \textbf{Restricted access} (authorized users upon request only): the complete paper generation agent with prompts, specific exploits, and large-scale fabrication scripts. All framework components require a responsible AI license with declaration of intended use and agreement not to fabricate academic content for distribution.

\paragraph{Deployment Recommendations.}
\textbf{For venues considering AI-assisted review:} (i) \textit{Mandatory disclosure} of AI usage to authors and reviewers; (ii) \textit{Score-flag coupling}—papers flagged with integrity concerns cannot receive acceptance without senior reviewer override; (iii) \textit{Audit trails} logging all model inputs, outputs, and integrity evidence; (iv) \textit{Human oversight} for all flagged submissions.
Automated integrity flags should be treated as triage signals for additional checking, not as accusations of misconduct; authors should retain a presumption of innocence and access to appeal or clarification mechanisms.
\textbf{For researchers using AI discovery systems:} Authors remain fully responsible for verifying that all content accurately reflects their actual experiments, implementations, and results. Fabricated or empirically unsupported claims, whether intentional or due to AI hallucination, constitute scientific misconduct regardless of the generation method.

\paragraph{Threshold Governance.}
Any deployment of integrity triage should be calibrated longitudinally, with explicit monitoring of false positives, false negatives, reviewer burden, and downstream appeal outcomes. Thresholds should be chosen to balance detection utility against unnecessary human effort and reputational harm.

\paragraph{Broader Impacts.}
AI-only publication loops threaten scientific epistemology. If fabrications become indistinguishable from genuine work, the foundation of scientific knowledge risks collapse. The path forward requires defense-in-depth across multiple layers: \textit{technical} (provenance verification, artifact validation), \textit{procedural} (integrity-aware scoring, human oversight), \textit{community} (post-publication review, whistleblower system), and \textit{cultural} (education on AI limitations, ethical guidelines).
We view this work as an \textit{early warning} system to catalyze robust defenses before these failure modes manifest at scale. Our findings demonstrate that current systems are not ready for AI-only research-the integrity of science depends on maintaining rigorous human evaluation as AI capabilities advance.

\section*{Acknowledgments}
This work is partially supported by the National Science Foundation (NSF) AI Institute for Agent-based Cyber Threat Intelligence and Operation (ACTION) under grant IIS 2229876. 

This work is supported in part by funds provided by the National Science Foundation, Department of Homeland Security, and IBM. 
Any opinions, findings, and conclusions or recommendations expressed in this material are those of the author(s) and do not necessarily reflect the views of the NSF or its federal agency and industry partners.

\bibliography{reference}
\appendix
\section{Supplementary}
\subsection{Stratified Sampling Procedure}\label{appx:sample-algo}
We implement the stratified sampling pipeline to construct the calibration corpus as follows. 

First, we partition the score space using bin edges $t_0 < \cdots < t_B$ to define score bins $B_b = [t_{b-1}, t_b)$ for $b = 1, \ldots, B$. 

For each bin--status combination $(b,c) \in \{1,\ldots,B\} \times \mathcal{C}_{\mathrm{stat}}$, we define:
\begin{align}
\begin{split}
    \mathcal{I}_{b,c} = \{i : h_i \in B_b, \sigma_i = c\}, \\
N_{b,c} = |\mathcal{I}_{b,c}|, \quad 
p_{b,c} = \frac{N_{b,c}}{N_\star},
\end{split}
\end{align}

where $N_\star = \sum_{b=1}^B \sum_{c \in \mathcal{C}_{\mathrm{stat}}} N_{b,c}$ is the total reference pool size.

Given a target calibration size $N_{\mathrm{cal}}$, we allocate samples to each cell using proportional allocation with the largest-remainder method:
\begin{align*}
    n'_{b,c} &= p_{b,c} N_{\mathrm{cal}}, \quad
n_{b,c} = \lfloor n'_{b,c} \rfloor \\
R &= N_{\mathrm{cal}} - \sum_{b,c} n_{b,c}.
\end{align*}
We then add one additional sample to the $R$ cells with the largest remainders $n'_{b,c} - \lfloor n'_{b,c} \rfloor$. 

Finally, we sample uniformly without replacement $\mathcal{S}_{b,c} \subseteq \mathcal{I}_{b,c}$ with $|\mathcal{S}_{b,c}| = n_{b,c}$ and construct:
\begin{align}
\begin{split}
\mathcal{D}_{\mathrm{cal}}  = \{(x_i, y_i^{\mathrm{hum}}, \sigma_i, h_i) : i \in \mathcal{S}\}, \\
\text{where}\quad \mathcal{S}  = \bigcup_{b=1}^B \bigcup_{c \in \mathcal{C}_{\mathrm{stat}}} \mathcal{S}_{b,c}.
\end{split}
\end{align}

This construction ensures that $\hat{p}^{\mathrm{cal}}_{b,c} = n_{b,c}/N_{\mathrm{cal}} \approx p_{b,c}$ for all $(b,c)$, preserving both score-bin and status marginals up to integer rounding.

\subsection{Error Analysis of Review Scoring}
\label{app:error-analysis}

Having defined our review aggregation mechanism, we now turn to a fundamental question: how reliable are the resulting scores and decisions? When we combine judgments from multiple reviewer agents, two sources of uncertainty arise. First, each reviewer introduces randomness—even when evaluating the same paper, a model may produce slightly different scores across runs. Second, our decision thresholds are estimated from finite calibration data and therefore subject to sampling error. 

We address these concerns by providing a rigorous error analysis that answers two questions:
\begin{itemize}[leftmargin=12pt]
  \item \textbf{Q1: How much does ensembling reduce randomness?} 
  Under independent reviewers, we give concentration bounds in Theorem 1 and Corollary 2 to show how tightly $s(x)$ clusters around its latent mean.
  \item \textbf{Q2: How reliable is a threshold picked from finite calibration data?}
  We give bounds on the acceptance-rate estimation error and the $0.5$-probability threshold with isotonic calibration in Propositions 1 and 2.
\end{itemize}
We also provide a Bayesian view that yields credible intervals for decision-making under uncertainty.

\paragraph{Assumptions.}
To make our analysis tractable, we impose two standard regularity conditions on the review process. For each model $m\in\mathcal{M}$, let $\mathbf{r}_m(x)\in\mathbb{R}^K$ denote the rubric vector and define the weighted consensus rubric $\bar{\mathbf{r}}(x)=\sum_{m}w_m\,\mathbf{r}_m(x)$. Let the latent mean be $\bar{\boldsymbol{\mu}}(x)=\sum_m w_m\,\mathbb{E}[\mathbf{r}_m(x)\mid x]$. We assume:
\begin{itemize}[leftmargin=10pt]
  \item (\textbf{Sub-Gaussian}) For each $m$, the centered rubric $\mathbf{z}_m(x):=\mathbf{r}_m(x)-\mathbb{E}[\mathbf{r}_m(x)\mid x]$ is \emph{vector sub-Gaussian}: for all $\mathbf{u}\in\mathbb{R}^K$, $\langle \mathbf{u},\mathbf{z}_m(x)\rangle$ is sub-Gaussian with proxy $\sqrt{\mathbf{u}^\top \Sigma_m \mathbf{u}}$. Moreover, $\{\mathbf{z}_m(x)\}_{m\in\mathcal{M}}$ are mutually independent.
  \item (\textbf{Lipschitz}) $\phi:\mathbb{R}^K\to\mathbb{R}$ is $L_\phi$-Lipschitz w.r.t.\ $\ell_2$: $|\phi(\mathbf{a})-\phi(\mathbf{b})|\le L_\phi\|\mathbf{a}-\mathbf{b}\|_2$.
\end{itemize}

These assumptions are natural in the peer-review setting. The sub-Gaussian property follows from the fact that venues always require bounded rubric scores, ensuring $r_{m,k}\in[a_k,b_k]$ and thus sub-Gaussianity via Hoeffding's lemma \citep{hoeffding1963}. 
The independence assumption reflects the standard practice that different reviewers evaluate papers independently without coordination. The Lipschitz condition is satisfied by common aggregation functions such as weighted averages ($\phi(\mathbf{a})=\mathbf{v}^\top \mathbf{a}$, $L_\phi=\|\mathbf{v}\|_2$) or selecting a single overall score ($L_\phi=1$).

With these assumptions in place, we define the latent target score $\mu_s(x):=\phi(\bar{\boldsymbol{\mu}}(x))$ as the score we would obtain if each reviewer's noise were averaged out. Under independence across reviewers, the aggregate vector noise has proxy matrix
\[
\Sigma_{\mathrm{vec}}(\mathbf{w})\;:=\;\sum_{m\in\mathcal{M}} w_m^2\,\Sigma_m \ \in\mathbb{R}^{K\times K},
\]
and we use the scalar variance proxy
\[
V_w \;:=\; \lambda_{\max}\!\big(\Sigma_{\mathrm{vec}}(\mathbf{w})\big).
\]

\paragraph{Frequentist concentration for ensemble scoring.}
We begin by quantifying how closely the observed ensemble score $s(x)$ tracks the latent mean $\mu_s(x)$. The following result shows that aggregating multiple independent reviewers yields exponentially tight concentration.

\noindent\textbf{Theorem 1 (Bernstein-McDiarmid concentration and margin bound).}
Under the assumptions above, let
$c_m := L_\phi\,w_m\sqrt{\sum_{k=1}^K(b_k-a_k)^2}$ and
$\sigma_w^2 := \mathrm{Var}[s(x)] \le L_\phi^2\sum_m w_m^2\,\lambda_{\max}(\Sigma_m)$,
with $c_{\max}:=\max_m c_m$.
Then for any $t>0$,
\begin{equation}\label{eq:concentration-one-sided}
    \Pr\!\big(s(x)-\mu_s(x)\ge t\big)
\;\le\;
\exp\!\Big(\!-\frac{t^2}{2\sigma_w^2+\frac{2}{3}c_{\max}t}\Big).
\end{equation}
Consequently, with $y^\star(x)=\mathbb{I}[\mu_s(x)\ge\tau]$ denoting the latent decision at threshold $\tau$ and
$\gamma(x)=|\mu_s(x)-\tau|$ denoting the margin,
\begin{equation}
\Pr\!\big(\hat{y}(x)\neq y^\star(x)\big)
\;\le\;
\exp\!\Big(\!-\frac{\gamma(x)^2}{2\sigma_w^2+\frac{2}{3}c_{\max}\gamma(x)}\Big).
\label{eq:margin-bound}
\end{equation}

\noindent\textbf{Corollary 1 (Variance-minimizing weights for linear aggregation).}
Suppose $\phi(\mathbf{a})=\mathbf{v}^\top \mathbf{a}$ is linear. 
Let $c_m:=\mathbf{v}^\top \Sigma_m \mathbf{v}$. Then $V_w=\sum_m w_m^2 c_m$ and among $\mathbf{w}\in\Delta^{M-1}$ 
the bound in \eqref{eq:margin-bound} is minimized by
\[
w_m^\star \ \propto\ \frac{1}{c_m}\ =\ \frac{1}{\mathbf{v}^\top \Sigma_m \mathbf{v}}\,,
\]
i.e., (diagonal) GLS/precision weighting in the projected variance.

\paragraph{Scalar-score simplification (overall assessment).}
The general vector-rubric framework of Theorem 1 applies when reviewers provide detailed multi-criterion scores. However, in many venues (e.g., ICLR/ICML), reviewers independently provide a single bounded \emph{overall assessment} that already aggregates rubric criteria internally. This special case admits a simpler analysis. Let each model output a scalar overall score $s_m(x)\in[a_m,b_m]$.

\noindent\textbf{Corollary 2 (Scalar overall-assessment bounds).}
If each reviewer outputs $s_m(x)\in[a,b]$ and $\phi$ is the identity,
then $\sigma_w^2=\sum_m w_m^2 \mathrm{Var}[s_m(x)]$ and $c_{\max}=\max_m w_m(b-a)$, hence
\begin{equation}\label{eq:scalar-margin-bound}
    \Pr\!\big(\hat{y}(x)\neq y^\star(x)\big)\;\le\; 
\exp\!\left(\!-\frac{\gamma(x)^2}{2\sigma_w^2+\frac{2}{3}c_{\max}\gamma(x)}\right).
\end{equation}
For uniform weights $w_m=1/M$ and identical per-review variance $\sigma^2$,
this simplifies to
\begin{equation}\label{eq:uniform-scalar-bound}
\Pr(\hat{y}\neq y^\star)\le
\exp\!\left(\!-\frac{M\,\gamma^2}{2\sigma^2+\frac{2}{3}(b-a)\gamma}\right),
\end{equation}
showing that both the variance term $\sigma^2/M$ and bounded-difference term $(b-a)/M$ scale as $1/M$.

\paragraph{Calibration error and threshold selection.}
The concentration results above assume a known threshold $\tau$. In practice, however, we must estimate $\tau$ from finite calibration data, introducing a second source of error. We now bound this calibration uncertainty. Let $\alpha(\tau):=\mathbb{P}_{x\sim\mathcal{D}_{\mathrm{cal}}}\big[s(x)\ge \tau\big]$ be the true acceptance rate at threshold $\tau$ on the calibration distribution, and let $\widehat{\alpha}_{\mathrm{cal}}(\tau)$ be its empirical counterpart (Section~\ref{sec:review-calibration}). The calibration set $\{x_i\}_{i=1}^{N_{\mathrm{cal}}}$ is treated as i.i.d.\ from $\mathcal{D}_{\mathrm{cal}}$.

\noindent\textbf{Proposition 1 (Calibration error bound).} For any $\delta\in(0,1)$, with probability at least $1-\delta$ over the draw of $\mathcal{D}_{\mathrm{cal}}$,
\begin{equation}
\sup_{\tau\in\mathbb{R}}\, \big|\,\widehat{\alpha}_{\mathrm{cal}}(\tau)-\alpha(\tau)\,\big| \;\le\; \sqrt{\tfrac{1}{2N_{\mathrm{cal}}}\log\tfrac{4}{\delta}}\,.
\label{eq:uniform-alpha}
\end{equation}
\emph{Proof sketch.} The class $\{\mathbb{I}[s\ge \tau]:\tau\in\mathbb{R}\}$ has VC dimension 1; apply the Dvoretzky–Kiefer–Wolfowitz (DKW) inequality with VC generalization to obtain \eqref{eq:uniform-alpha}. \hfill$\square$

This uniform bound controls the acceptance-rate error across \emph{all} thresholds simultaneously. For the rate-matching threshold $\tau_{\mathrm{rate}}$ (defined to match the venue's historical acceptance rate $\alpha^\star$), we therefore have $|\widehat{\alpha}_{\mathrm{cal}}(\tau_{\mathrm{rate}})-\alpha^\star|\le \sqrt{\frac{1}{2N_{\mathrm{cal}}}\log\frac{4}{\delta}}$. If $\alpha(\tau)$ is strictly decreasing with slope bounded away from zero near $\tau_{\mathrm{rate}}$, this acceptance-rate error translates into a correspondingly small threshold error.

For the probability-consistency threshold $\tau_{0.5}$, the analysis is more delicate because we must estimate the conditional acceptance probability $\pi(t)=\mathbb{P}(y^{\mathrm{hum}}=1\mid s(x)\ge t)$ and then invert it. We employ isotonic regression to ensure monotonicity, and the following result bounds the resulting threshold error.

\noindent\textbf{Proposition 2 (Bound for $\tau_{0.5}$ with isotonic calibration).}
Define the generalized inverses $\tau_{0.5}=\inf\{t:\pi(t)\ge 1/2\}$ and $\hat{\tau}_{0.5}=\inf\{t:\hat{\pi}(t)\ge 1/2\}$. Suppose $\sup_t |\hat{\pi}(t)-\pi(t)|\le \varepsilon_\pi$ and $\pi$ has no flat region wider than $\Delta$ around $\tau_{0.5}$ and let $c_{\min}$ be the minimal right-slope of $\pi$ at $\tau_{0.5}$. Then
\begin{equation}
|\,\hat{\tau}_{0.5}-\tau_{0.5}\,| \;\le\; \min\{\Delta,\ \varepsilon_\pi / c_{\min}\}.
\label{eq:tau05-bound}
\end{equation}
\emph{Proof sketch.} Since $\pi$ is monotone with right-slope $c_{\min}$, $\pi(\tau_{0.5}+h)\ge \tfrac{1}{2}+c_{\min}h$ and $\pi(\tau_{0.5}-h)\le \tfrac{1}{2}-c_{\min}h$ for $0<h\le\Delta$; with $\sup_t|\hat{\pi}-\pi|\le\varepsilon_\pi$, choosing $h=\min\{\Delta,\varepsilon_\pi/c_{\min}\}$ yields $\hat{\pi}(\tau_{0.5}+h)\ge\tfrac{1}{2}$ and $\hat{\pi}(\tau_{0.5}-h)\le\tfrac{1}{2}$, hence $|\hat{\tau}_{0.5}-\tau_{0.5}|\le h$. \hfill$\square$

\begin{figure*}[t]
\centering
\includegraphics[width=\textwidth]{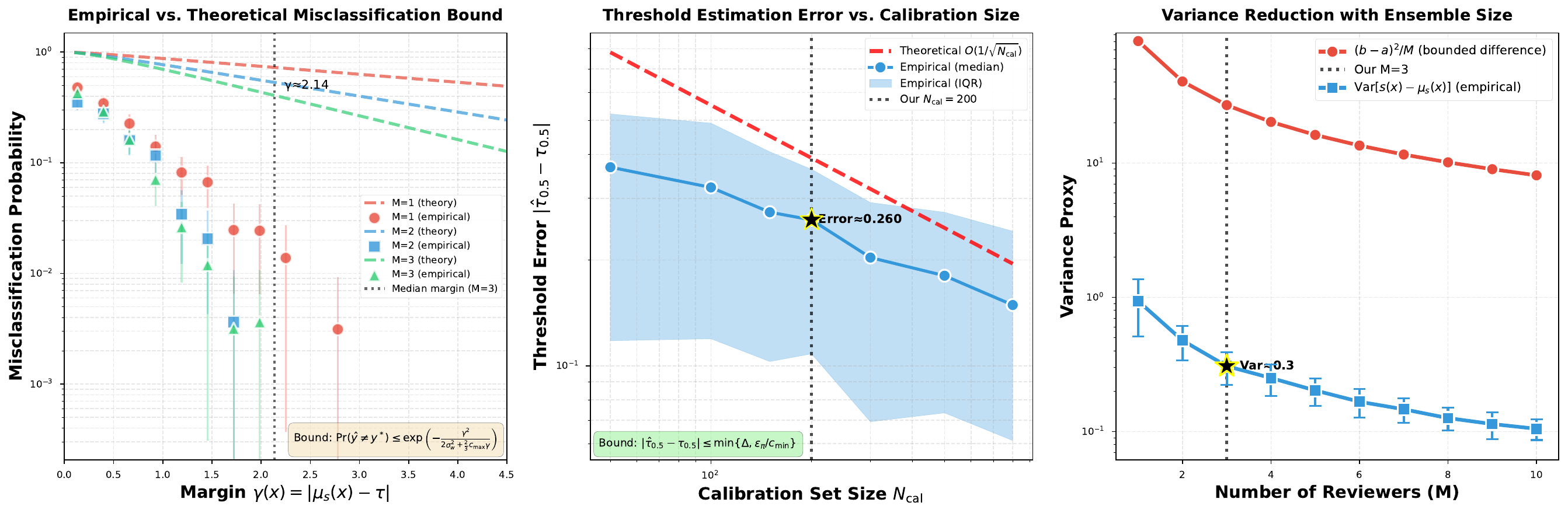}
\caption{\textbf{Empirical validation of error analysis bounds.} 
\textbf{Left:} Misclassification probability vs.\ margin $\gamma(x)$ for $M=1,2,3$ reviewers. Empirical rates (points with error bars) fall below theoretical bounds (dashed lines), confirming Eq.~\eqref{eq:margin-bound}. 
\textbf{Middle:} Threshold estimation error vs.\ calibration set size $N_{\mathrm{cal}}$. The blue curve follows the theoretical $O(1/\sqrt{N_{\mathrm{cal}}})$ decay (red dashed); our $N_{\mathrm{cal}}=200$ (star) yields error $\approx 0.26$, validating Proposition~1. 
\textbf{Right:} Variance reduction with ensemble size (log scale). Both the empirical noise variance $\mathrm{Var}[s(x)-\mu_s(x)]$ (blue squares) and the bounded-difference proxy $(b-a)^2/M$ (red circles) decrease as $1/M$, demonstrating that increasing from $M=1$ to $M=3$ reviewers reduces both quantities by approximately $3\times$—confirming Theorem~1 and Corollary~2.}
\label{fig:error-analysis-validation-appendix}
\end{figure*}

\paragraph{Bayesian credible decisions.}
The frequentist bounds above provide worst-case guarantees but do not directly yield decision rules for individual papers. We complement this analysis with a Bayesian perspective that provides paper-specific uncertainty quantification. Assume $s_m(x)\mid \mu(x)\sim \mathcal{N}(\mu(x),\sigma_m^2)$ independently across $m$ and 
$\mu(x)\sim \mathcal{N}(\mu_0,\tau_0^2)$. Then the posterior is Gaussian with precision and mean given by
\begin{align}
    \tau_n^{-2}& =\;\tau_0^{-2}\;+\;\sum_{m}\sigma_m^{-2} \; ,
\label{eq:posterior-precision}\\
\mu_n &=\;\tau_n^2\!\left(\mu_0\tau_0^{-2}\;+\;\sum_m \sigma_m^{-2}\,s_m\right)\;.
\label{eq:posterior-mean}
\end{align}

For any threshold $\tau$, the posterior decision probability is $\mathbb{P}(\mu(x)\ge \tau\mid\{s_m\})=1-\Phi((\tau-\mu_n)/\tau_n)$. A $1-\alpha$ credible decision is robust (i.e., the credible interval for $\mu(x)$ does not straddle the threshold) whenever $|\tau-\mu_n|\ge z_{1-\alpha/2}\,\tau_n$. 

This Bayesian framework also provides a principled rule for soliciting additional reviews. If the current decision is ambiguous ($|\tau-\mu_n|< z_{1-\alpha/2}\,\tau_n$) and a candidate reviewer with variance $\sigma_{\text{new}}^2$ would resolve the ambiguity in expectation—that is,
\[
|\tau-\mu_n|\ \ge\ z_{1-\alpha/2}\,\tau_{n+1},
\quad 
\tau_{n+1}^{-2}=\tau_n^{-2}+\sigma_{\text{new}}^{-2},
\]
then the additional review is worthwhile; otherwise, the expected uncertainty reduction is insufficient to justify the cost. This credible-interval framework thus enables both probability-of-acceptance decisions and adaptive review allocation.
\paragraph{Empirical validation.} 
To validate our theoretical bounds, we conduct synthetic experiments that simulate the review aggregation process under controlled conditions. We generate $n=5{,}000$ synthetic papers with known latent quality scores, each reviewed by $M\in\{1,2,3\}$ independent models producing noisy scalar assessments in $[1,10]$. For each configuration, we compute: (i) empirical misclassification rates as a function of margin $\gamma(x)$ and compare against the bound in \eqref{eq:margin-bound}; (ii) threshold estimation error $|\hat{\tau}_{0.5}-\tau_{0.5}|$ for varying calibration set sizes $N_{\mathrm{cal}}\in\{50,\ldots,800\}$ via bootstrap with isotonic regression; (iii) empirical noise variance $\mathrm{Var}[s(x)-\mu_s(x)]$ and the bounded-difference proxy $(b-a)^2/M$ as functions of ensemble size $M$.

Figure~\ref{fig:error-analysis-validation-appendix} presents the results. The left panel confirms that empirical misclassification rates fall well below the theoretical bound across all margins and ensemble sizes, with clear separation between $M=1,2,3$ demonstrating the benefit of aggregation. The middle panel shows threshold error decreasing as $O(1/\sqrt{N_{\mathrm{cal}}})$ as predicted by Proposition 1, with our choice of $N_{\mathrm{cal}}=200$ (marked by the star) yielding error $\approx 0.26$ at the operating point. The right panel demonstrates how increasing the number of reviewers reduces both sources of uncertainty: the empirical noise variance $\mathrm{Var}[s(x)-\mu_s(x)]$ (blue squares) and the bounded-difference proxy $(b-a)^2/M$ (red circles) both decrease as $1/M$. Increasing from $M=1$ to $M=3$ reviewers reduces both quantities by approximately $3\times$—confirming that recruiting additional independent reviewers substantially improves decision reliability. These empirical results validate that our bounds correctly characterize the system's behavior.

\paragraph{Practical implications.}
Taken together, the error analysis in this section yields three actionable recommendations for deploying agentic review systems:
\begin{itemize}[leftmargin=10pt]
\item[\textbf{(i)}] \textbf{Aggregate intelligently.} Keep the variance proxy $V_w$ small by recruiting independent reviewers and using variance-aware weighting (e.g., Corollary 1).
\item[\textbf{(ii)}] \textbf{Handle borderline cases carefully.} When the margin $|s(x)-\tau|$ is small, use Bayesian credible intervals to assess decision confidence and determine whether additional reviews are needed.
\item[\textbf{(iii)}] \textbf{Calibrate sufficiently.} Choose $N_{\mathrm{cal}}$ large enough so that the DKW deviation in \eqref{eq:uniform-alpha} is negligible at the target confidence level (e.g., $N_{\mathrm{cal}}\ge 200$ for $\delta=0.05$ yields uniform error $\lesssim 0.11$).
\end{itemize}

\section{Seed Topic List} \label{appx:seed-topic-list}
We use \texttt{GPT-5} to generate $25$ seed topics aligned with the ICLR submission calibration corpus, covering AI, ML, CV, NLP, robotics, systems, and security:
\begin{itemize}[leftmargin=8pt, itemsep=2pt, parsep=0pt]
  \item Self-consistent diffusion models that satisfy counterfactual causal constraints.
  \item Open-world continual evaluation via synthetic task evolution for multimodal LLMs.
  \item Mechanistic interpretability of Mixture-of-Experts routing as a cooperative game.
  \item Certified robustness for retrieval-augmented generation under adversarial knowledge bases.
  \item Neural field memory: spatially grounded long-horizon memory for vision-language agents.
  \item Program-of-Thought VLMs with verifiable tool-use and executable intermediate graphs.
  \item On-device nano-LLMs co-designed with NPU schedulers for sub-1W edge inference.
  \item Causal video generation: 4D text-to-video with physics-invariant latent constraints.
  \item Self-curating agents: autonomous dataset construction with legal/ethical compliance proofs.
  \item Safety proofs for multi-agent LLM protocols under Byzantine participants.
  \item Open-vocabulary 3D segmentation with Gaussian splats and generative object priors.
  \item Unlearning at scale: certified removal of concepts from multimodal foundation models.
  \item Temporal reasoning benchmarks for VLMs built from parametric CAD + differentiable physics.
  \item Federated reinforcement learning with privacy-preserving credit assignment.
  \item Energetically aligned training: minimizing carbon under fixed accuracy via differentiable scheduling.
  \item Watermarking as cryptographic dialogue: interactive proofs to verify AI-generated media.
  \item Neurosymbolic chart-to-code: parsing scientific plots into executable analysis programs.
  \item Robust long-form instruction following via adversarial curriculum from self-play reviewers.
  \item World-model rewrites: editing factual and procedural knowledge in LLMs with locality guarantees.
  \item Haptic-vision-language models for household manipulation with uncertainty-aware plans.
  \item Compositional diffusion: plug-and-play constraints for safety, style, and identity preservation.
  \item Reasoning-first pretraining: supervising latent chains over captions, code, and proofs.
  \item Open-set alignment: detecting and mitigating unseen harms in generative agents at test time.
  \item Graph-grounded RAG: joint learning of knowledge graphs and retrievers for verifiable answers.
  \item RouteBench: measuring strategic routing, tool selection, and delegation in multi-agent LLM systems.
\end{itemize}
\end{document}